\documentclass[10pt,conference]{IEEEtran}
\IEEEoverridecommandlockouts
\usepackage{cite}
\usepackage{amsmath,amssymb,amsfonts}
\usepackage{algorithmic}
\usepackage{graphicx}
\usepackage{textcomp}
\usepackage{xcolor}
\usepackage[tikz]{mdframed} 
\usepackage[textsize=small]{todonotes}
\usepackage{booktabs}
\usepackage[hidelinks]{hyperref}
\usepackage{array}
\usepackage{multirow}

\newcommand\SequentialCode{\textsc{Sequ+Code+Desc}}
\newcommand\SingleCode{\textsc{Code+Desc}}
\newcommand\SequentialNoCode{\textsc{Sequ+Desc}}
\newcommand\SingleNoCode{\textsc{Desc}}
\newcommand\SeqCodeNoDesc{\textsc{Sequ+Code}}
\newcommand\SngCodeNoDesc{\textsc{Code}}
\newcommand\SeqNoCodeNoDesc{\textsc{Sequ}}
\newcommand\NoInformation{\textsc{NoInfo}}

\newcommand\researchquestion[2]{
    \vspace{0.6em}
    \begin{mdframed}[
    	linewidth=1pt,
    	leftmargin=0pt,
    	topline=false,
    	rightline=false,
    	bottomline=false,
    	linecolor=gray!50]
    \begin{tabular}{@{}l m{0.89\textwidth}}
    	\textbf{{#1}} & {#2} \\
    \end{tabular}
    \end{mdframed}
    \vspace{0.3em}
}

\newcommand\RQAnswer[2]{
	\noindent 
	\begin{mdframed}[
		linewidth=1pt,
		leftmargin=0pt,
		backgroundcolor=gray!10,
		linecolor=gray!10]
		\begin{tabular}{@{}l@{\hspace{6pt}} m{0.91\textwidth}}
			\textbf{{#1}} & {#2} \\
		\end{tabular}
	\end{mdframed}
}

\newcommand\PromptBox[1]{%
  \par
  \vspace{\topsep}
  \noindent
  \begingroup
  \setlength{\fboxsep}{6pt}%
  \colorbox{gray!20}{%
    \begin{minipage}{\dimexpr\linewidth-2\fboxsep\relax}
        \footnotesize
        \texttt{#1}      
    \end{minipage}
  }%
  \endgroup
  \par
  \vspace{\topsep}%
}

\begin{document}

\title{Predicting Program Comprehension with Foundation Models of Human Cognition}

\author{\IEEEauthorblockN{Yannick Lehmen, Marvin Wyrich, Anna-Maria Maurer, Norman Peitek, Sven Apel}
\IEEEauthorblockA{\textit{Saarland Informatics Campus, Saarland University} \\
Saarbrücken, Germany}
}

\maketitle

\begin{abstract}
Software engineering depends on the ability of developers to understand code, yet predicting how they do so remains an open challenge despite decades of research.
Existing approaches rely either on simplified proxy measures that limit accuracy or on non-trivial measurements requiring elaborate experimental setups that are difficult to scale and apply in practice.
In contrast, recent work in psychology suggests an alternative perspective: Instead of modeling task-specific phenomena directly, human behavior can be captured through general cognitive regularities learned from large-scale behavioral data.
This idea treats complex human behavior as the observable outcome of underlying cognitive processes that manifest consistently across tasks and domains.

In this paper, we explore this perspective in the context of program comprehension.
We evaluate Centaur, a foundation model trained on 160 general psychological experiments, on 9 previously published program-comprehension studies.
We assess how well its predicted response distributions align with human response data and compare Centaur's performance to its base model, Llama~3.1.
To better understand the source of its performance, we subsequently conduct ablation studies to isolate the contribution of different sources of information, such as the code artifacts, task-related context, and prior trials and participant responses.

In a nutshell, we find that Centaur more closely aligns with human response patterns than its base model, is significantly less reliant on information from prior trials and responses, and benefits more from task-related information.
These findings suggest that behavioral patterns learned from general psychological data can transfer to complex software engineering tasks such as program comprehension.
More broadly, they point toward foundation models of human cognition as a basis for modeling developer behavior in software engineering, opening a pathway toward a unified, data-driven perspective on human-centered software engineering grounded in cognitive science.
\end{abstract}


\section{Introduction}
\label{sec:intro}

Understanding how developers comprehend code is central to improving software quality and developer productivity~\cite{Xia:2018:Comprehension, Cheng:2022:Productivity, Gilmore:1991:Debugging}.
A wide range of approaches has been proposed to assess and predict the process of program comprehension, including controlled experiments with developers~\cite{Wyrich:2023:40Years}, proxy measures such as complexity metrics~\cite{Scalabrino:2019:Metrics}, and neurocognitive methods that measure brain activity during programming tasks~\cite{Siegmund:2020:Crazy}.
These approaches reflect different operationalizations of program comprehension, each capturing particular aspects of how humans engage with code.

Yet, despite this diversity of approaches, reliably predicting how developers will respond in program comprehension tasks remains an open challenge: Controlled experiments with programmers provide precise insights but do not scale beyond specific experimental settings, while automated proxy measurements such as complexity metrics offer scalability but show only weak correlations with observed human responses~\cite{Scalabrino:2019:Metrics, Peitek:2021:Metrics, Munoz:2020:CognitiveComplexity}.
Even more direct approaches, such as neurocognitive methods that measure brain activity during programming tasks, provide rich signals but are expensive, difficult to generalize, and limited in scope~\cite{Sharafi:2021:Toward,Mcdowell:2013:Neuroimaging}.
Across all these approaches, a common limitation is that they do not capture transferable regularities in how humans respond, making it difficult to build predictive models of program comprehension that generalize beyond specific study designs and individuals.

We look towards psychology for a new perspective: Faced with issues of explainable models for human behavior often being ill-suited to predictive modeling and generalization beyond single studies, the field of psychology has seen an increase in the use of black-box machine learning models trained on behavioral data to predict how participants respond to certain tasks~\cite{Pargent:2023:MLPsych, Yarkoni:2017:Prediction, Shaw:2022:Behavior}.
These models do not provide explicit explanations of underlying cognitive processes, but they achieve a remarkably strong alignment with observed response patterns of humans, offering a complementary way to study human behavior.

Recent work by Binz et al.~\cite{Binz:2025:Centaur} introduced \emph{Centaur}, a large language foundation model fine-tuned on data from 160 psychological experiments with the goal of reproducing human behavior.
Unlike approaches that rely on task-specific measurement designs or structured models of behavior, Centaur learns directly from large-scale behavioral data, allowing it to capture variability in how humans respond across different experimental settings.
Experiments by Binz et al.~\cite{Binz:2025:Centaur} show that Centaur more closely aligns with human data than both its base model, Llama~3.1, and established interpretable models of human behavior across diverse psychological experiments.
Moreover, follow-up work demonstrates that Centaur can be used to systematically improve such interpretable models, identifying gaps and guiding revisions that increase their predictive accuracy~\cite{Binz:2025:Automated}.
These results suggest that training on behavioral data enables models to capture generalizable regularities in how humans respond to tasks.
However, it is unclear whether such regularities extend to substantially more complex and structured tasks in different domains, such as program comprehension.
This motivates us to investigate whether foundation models of human cognition can exhibit similar alignment in such settings.
While the training data underlying Centaur cover only relatively simple experimental paradigms, such as n-back memory tasks~\cite{Binz:2025:Centaur}, it has been shown to capture fundamental regularities of human cognition and decision-making.
If these regularities are sufficiently general, they may transfer to more complex domains, enabling prediction of human responses in program comprehension tasks despite their richer structure and semantic complexity.

In this paper, we evaluate this hypothesis by studying whether a foundation model of human cognition can predict human responses in program comprehension tasks.
Specifically, we analyze how well Centaur aligns with human response distributions across 9 program comprehension studies drawn from prior work.
We compare its performance to its base model, Llama~3.1~\cite{Llama3}, to isolate the effect of fine-tuning on behavioral data, and conduct ablation studies to examine whether improvements are driven by task-relevant information (such as task descriptions or code) rather than generic statistical patterns in human response data (i.e., response biases).

Our results suggest that models trained on psychological experiment data provide a transferable basis for predicting human responses in a program comprehension setting, indicating that basic cognitive regularities may generalize across task complexity and domain boundaries.
In our ablation studies, we find that compared to its base model, Centaur benefits more from the presence of task-relevant information and less from the presence of prior responses, indicating that the transfer from psychological experiments to program comprehension is not based on response biases but instead on a better modeling of program comprehension.

At a practical level, our results demonstrate, that such models can be used to approximate how developers respond in program comprehension tasks, offering a new, data-driven way to study and anticipate human behavior in this domain.
Beyond this immediate implication, our findings point toward two broader perspectives.
First, they suggest the potential for a unified, data-driven framework for modeling human-centered activities in software engineering grounded in cognitive science, where behavioral data, including data collected in psychological experiments, may serve as a common foundation across a range of tasks, such as debugging, reviewing pull requests, or estimating development effort.
Second, the ability of Centaur to align with human responses in program comprehension tasks indicates that this task is shaped by general cognitive regularities captured in the underlying behavioral data.
This opens up new opportunities to study program comprehension as an instance of general cognitive regularities, providing a novel lens on how developers understand code.

\section{Background \& Related Work}

In the following, we provide some background and summarize prior work related to our study. 
We review three key areas: (1)~theories of program comprehension, (2)~measuring and modeling program comprehension, and (3)~predictive modeling of general human behavior with black-box approaches, such as Centaur.

\subsection{Cognitive Theories of Program Comprehension}

Program comprehension refers to a developer's intentional act and degree of accomplishment in inferring the meaning of source code~\cite{Wyrich:2023:CC}. 
Due to its importance to everyday activities of developers~\cite{Minelli2015, Xia:2018:Comprehension}, it has long been of interest to researchers~\cite{Wyrich:2023:40Years}.
Classical theories distinguish between bottom-up and top-down comprehension.
In bottom-up comprehension, developers build their understanding from individual syntactic elements, gradually integrating their meaning toward a higher-level semantic understanding~\cite{Pennington1987}.
In contrast, in the more efficient~\cite{Siegmund2017} top-down comprehension, developers use knowledge and experience in a hypothesis-driven comprehension process, where hypotheses about the program are generated, tested, and refined until their relationship to the code becomes apparent~\cite{Brooks1978, Brooks1983}.
In practice, research has found that developers integrate both strategies, switching between bottom-up and top-down comprehension depending on the task and context~\cite{Mayrhauser1997, Littman1987}. 
Further, program comprehension is influenced by substantial individual differences between developers~\cite{Navarro2025, McConnell2011}.
Despite decades of research, it remains unclear how experience, expertise, and knowledge play a role in efficient and effective program comprehension~\cite{Peitek2022}.

A key aspect to understanding program comprehension is how developers process the vast information of source code.
Prior research has identified several cognitive mechanisms to play an important role.
For example, chunking refers to grouping information together into useful units, which not only is a critical process in program comprehension~\cite{Shneiderman1979}, but also of general use for humans related to their working memory and attention.
Neuro-cognitive studies on program comprehension have revealed an involvement of natural language processing~\cite{Siegmund:2014:fMRI, Floyd:2017:fMRI}, formal math and logic~\cite{Liu2020}, multi-demand networks~\cite{Ivanova:2020:fMRI}, and more~\cite{Castelhano2018}.
Together, these findings highlight that program comprehension is combining fundamental cognitive processes, which supports the idea that cognitive regularities from non-programming tasks may transfer to program comprehension.
Yet, while these theories provide conceptual models, they are difficult to operationalize for predicting how human developers will behave across a wide range of program-comprehension tasks.

\subsection{Measuring and Modeling Program Comprehension}

Since program comprehension is inherently difficult to observe directly~\cite{Siegmund2016} with many confounding factors~\cite{Siegmund:2015:Confounding}, research and practice has adopted a wide range of approaches to measure program comprehension.

Controlled experiments are the primary method for studying program comprehension, where researchers gain insights from developers based on self-reflection~\cite{Kosti2018}, think-aloud protocols~\cite{Pennington1987, Shaft1995}, or behavioral measures~\cite{Soloway1984}, such as correctness and response times.
These studies are limited in scale and context, making it difficult to generalize their findings across different tasks and settings.

As a complementary approach, researchers have proposed and investigated proxy measures, such as code complexity metrics, to predict program comprehension based on static code features~\cite{Beyer2010, SonarSource2018:CognitiveComplexity}.
These measures can be computed automatically and therefore scale to industrial settings.
However, they simplify the multi-dimensionality of source code, individual developer characteristics, and program comprehension into a narrow set of proxies, often expressed as a single number.
Moreover, complexity metrics often operationalize one specific aspect of program comprehension for a specific task and settings.
So while complexity metrics provide a scalable and cheap measurement, empirical studies have consistently shown that they exhibit weak relationships with human behavior--at best~\cite{Scalabrino:2019:Metrics, Peitek:2021:Metrics, Hao2023}.

More recently, researchers have been transferring methods from neuroscience to study program comprehension, such as eye tracking~\cite{Obaidellah2018}, electroencephalogram (EEG)~\cite{Gonccales2021}, or functional magnetic resonance imaging (fMRI)~\cite{Gonccales2021}, with some position papers promisingly outlining the possibilities of gaining more objective insights~\cite{Fakhoury:2018:Moving} or even an entirely new, neurocognitive perspective on program comprehension~\cite{Peitek:2018:Neuro, Siegmund:2020:Crazy}.
Yet, while neurocognitive methods provide experiment-specifically rich insights, they suffer from the same limitations as controlled experiments and additionally are time-consuming and expensive~\cite{Floyd:2017:fMRI}. 

To overcome the limitations of individual experiments, researchers have begun to explicitly model and predict developer behavior.
These approaches include probabilistic statistical models to predict code searching~\cite{McMillan2011} or gaze paths~\cite{Barthelemy2023}, machine learning to predict developer states from physiological data~\cite{Fritz2014}, or cognitive architectures, such as ACT-R~\cite{Ritter2019}, to predict human behavior~\cite{Hansen2012, ClosheimThesis}.

In a nutshell, existing approaches provide valuable insights into how developers understand source code, but they typically trade-off scalability, ecological validity, predictive accuracy, and generalizability.
As a result, we still lack a predictive model that can scale across diverse settings, account for inherent individual particularities, and accurately predict how developers respond to program-comprehension tasks.

\subsection{Predictive Modeling of Human Behavior}

To address similar challenges, researchers in psychology shifted from explicit explanatory models toward prediction-oriented models.
Rather than attempting to model cognitive processes with interpretable cognitive architectures, researchers use machine learning trained on large amounts of observable behavioral data.
A key idea is that this approach can internalize transferable regularities of human cognition and behavior~\cite{Pargent:2023:MLPsych, Yarkoni:2017:Prediction, Shaw:2022:Behavior}.
By learning these regularities, such predictive models can accurately predict human behavior beyond the specific tasks and context of the training data.
Although this limits the interpretability of the resulting model, it promises to provide a substantially better and more generalizable predictive framework~\cite{Yarkoni:2017:Prediction}.

With this perspective, Binz et al. introduced \emph{Centaur}, a foundation model of human cognition~\cite{Binz:2025:Centaur}.
To this end, they fine-tuned the large language model Llama~3.1~\cite{Llama3} with behavioral data from a diverse set of 160~psychological experiments.
Their dataset consisted of lab experiments dealing with many different cognitive processes, such as multi-armed bandit experiments, which investigate exploration-exploitation trade-offs in probabilistic games, or n-back tasks investigating working memory. 
Their results show that Centaur exhibits a stronger alignment with human behavior compared to both its base model Llama~3.1 and existing cognitive models across a range of psychological tasks, including tasks not present in its training data.
This demonstrated that Centaur does not only learn human behavior for specific tasks, but is able to transfer behavioral regularities across tasks. 
Furthermore, Binz et al. subsequently showed how Centaur can be used to construct interpretable models by iteratively refining an interpretable model of a specific task until it approaches the predictive performance of Centaur~\cite{Binz:2025:Automated}.
In this way, foundation models of human cognition can be used both to predict human behavior across a wide variety of experimental tasks by transferring behavioral regularities across tasks and pave the way for more accurate explanatory models.

\subsection{Research Gap}

Despite decades of research, we still lack scalable and accurate predictive models on human behavior for tasks in program comprehension.
In contrast, recent work in psychology demonstrates that data-driven models can successfully capture behavioral regularities.
In this paper, we address the open question whether foundation models of human cognition trained on behavioral data from psychological tasks, such as Centaur, can transfer insights to complex tasks in software engineering.
Program comprehension shares several facets with general cognitive tasks, such as working memory, attention, natural language and semantic processing, which suggests that a transfer might be possible, but no prior work has systematically investigated this possibility.

\section{Methodology}

\begin{figure*}[ht!]
    \centering
    \includegraphics[width=1\linewidth]{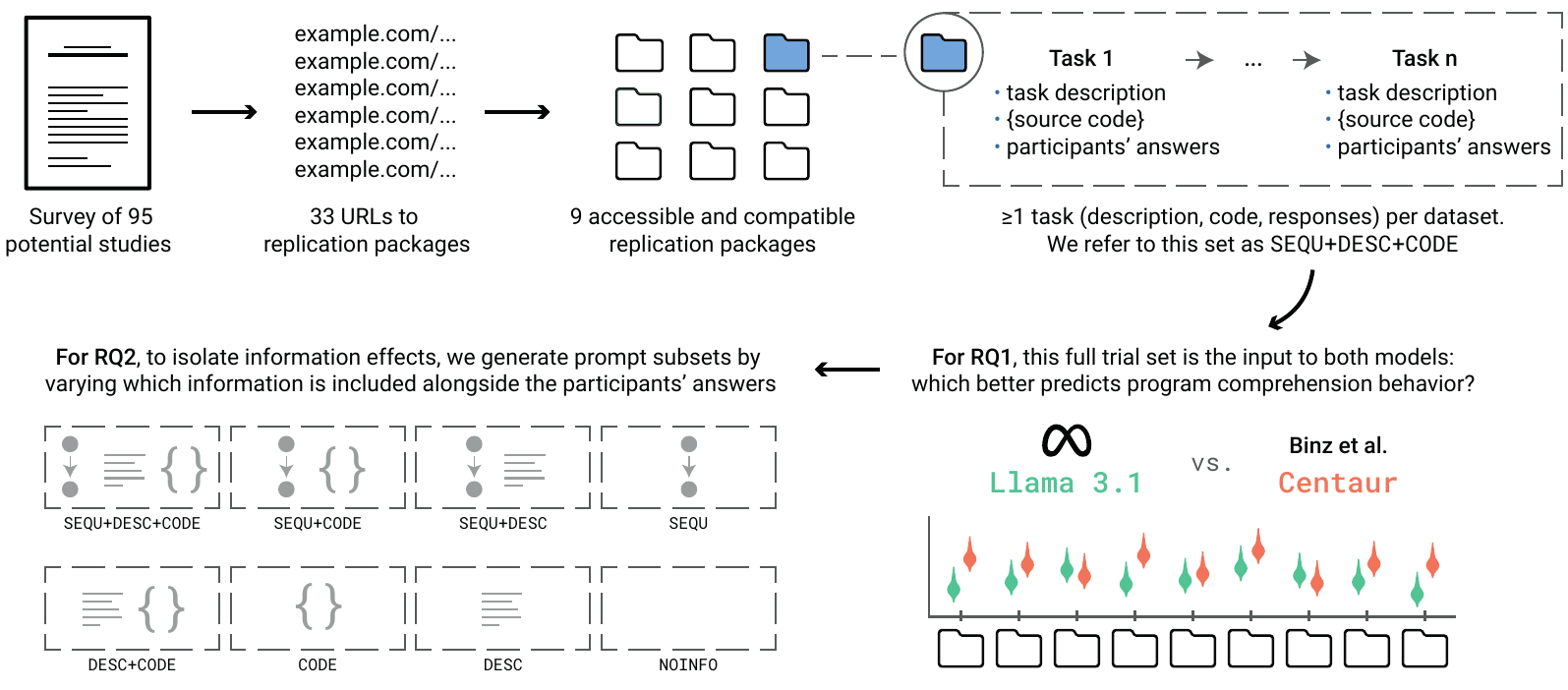}
    \caption{Overview of the evaluation pipeline, from dataset construction to model evaluation. We compare Centaur and Llama~3.1 on predicting human response distributions (RQ\textsubscript{1}) and use input ablations to analyze sources of performance differences (RQ\textsubscript{2}).}
    \label{fig:method_overview_full}
\end{figure*}

Our goal is to investigate whether foundation models of human cognition trained on psychological data can accurately predict developer behavior in program comprehension. 
For this purpose, we compare Centaur with its base model, Llama~3.1~70B (hereafter referred to as Llama), in their ability to predict human responses across datasets from prior experiments.
Evaluating both models under identical conditions allows us to isolate the effect of fine-tuning on human behavioral data.
In that vein, we pose our first research question: 
\researchquestion{RQ\textsubscript{1}}{To what extent does fine-tuning Llama on behavioral data from psychological experiments affect the prediction of human response distributions in program-comprehension tasks?}

While RQ\textsubscript{1} allows us to isolate the effect that fine-tuning on behavioral data has on the performance of Centaur, it cannot pinpoint the source of this improvement. 
In particular, it remains unclear whether possible improvements stem from generic statistical patterns in human responses (i.e., response biases) or from information specific to program comprehension tasks (e.g., code semantics, task descriptions).
To disentangle these effects, we conduct ablations in which we systematically vary the information available to the model. 
In that sense, we pose our second research question:

\researchquestion{RQ\textsubscript{2}}{To what extent are performance differences between Centaur and Llama explained by task-relevant information?}

Together, RQ\textsubscript{1} and RQ\textsubscript{2} allow us to quantify the impact fine-tuning on behavioral data has on predicting human response distributions in program comprehension and identify how different sources of information affect these potential improvements.
\autoref{fig:method_overview_full} provides a schematic overview of our research methodology, which we explain next in more detail.

\subsection{Dataset Collection and Selection}

We constructed our evaluation dataset by systematically filtering prior program-comprehension studies to retain only those with accessible, well-structured behavioral data suitable for our analysis.
As a starting point, we used the comprehensive literature review on program-comprehension studies by Wyrich et al.~\cite{Wyrich:2023:40Years}, which covers 95 articles. To identify viable datasets, we applied the following steps:

\begin{enumerate}
    \item \textbf{Initial filtering.} Since we require access to the data collected in studies, we began by excluding all articles that did not provide a link to a replication package. To identify such links, we searched for a predefined set of keywords\footnote{A full list of keywords is provided in our replication package~\cite{zenodo:dataset}.} and manually discarded false positives. This step excluded 62 papers, leaving 33 papers.
    \item \textbf{Study-level consolidation.} We separated the papers into studies, yielding 33 studies, as one paper dealt with two studies and two papers used data from the same study.
    \item \textbf{Task suitability.} We excluded all studies that were unfit for our evaluation approach. This included studies that could not be meaningfully translated into natural language (e.g., class diagram tasks), those with unconstrained response formats (e.g., lengthy free-text responses), and those unfit for generating prompts (e.g., because a full prompt would exceed the context window used in the evaluation). When only specific tasks within a study were unsuitable, we excluded those tasks while retaining the rest of the study. This step excluded 14 studies, leaving 19 studies.
    \item \textbf{Accessibility of replication packages.} We verified whether the replication packages linked in the studies' papers were accessible. In seven cases, access was not possible, so we contacted the authors. This was successful for three datasets, while four remained inaccessible and were excluded. This resulted in 15 studies.
    \item \textbf{Completeness of behavioral data.} Finally, we ensured that each dataset contained all necessary information: the full set of tasks, their presentation order, and the responses each participant gave for each task. We excluded six studies that did not meet these requirements, yielding a final set of nine studies (see~\autoref{tab:IncludedExperiments} for an overview).
\end{enumerate}

\begin{table*}
    \centering
    \caption{Overview of included experiments with participants, trials, and task/response summaries. Reported counts reflect the data used in our study and may differ from the original datasets; deviations are documented in our replication package.}
    \label{tab:IncludedExperiments}
    \begin{tabular}{lp{30mm}lrrrp{22mm}rl}
        \toprule
         Study & Topic & Pro.  & \# of  & \# of Unique& \# of & Task& \# of & Response \\
          &  &  Language & Part. & Snippets/Tasks& Trials  & &Responses& Options \\
         \midrule
         S1~\cite{Ajami:2019:complexity}  & Code complexity & JavaScript & 208 & 40 & 2067& Output&2067 & Text\\
         S2~\cite{Bauer:2019:Indentation}  & Indentations & Java & 22 & 16 & 88& Output& 88 & Text\\
         \multirow[t]{2}{*}{S3~\cite{Borstler:2016:chains}}  & \multirow[t]{2}{30mm}{Method chains and comments} & \multirow[t]{2}{*}{Java} & \multirow[t]{2}{*}{104} & \multirow[t]{2}{*}{30} & \multirow[t]{2}{*}{506} & Readability & 1012 &Likert Scale  \\
         & & & & &  &Cloze test&1586 & Text\\
         S4~\cite{Buse:2010:metricreadability}  & Code readability & Java & 121 & 100 & 12100 & Readability&12100& Likert Scale\\
         S5~\cite{Gopstein:2017:atoms}  & Atoms of confusion & C & 73 & 126 & 6069 & Output&6069 & Text\\
         S6~\cite{Gopstein:2017:atoms}  & Atoms of confusion & C & 42 & 8 & 168 & Output&498 & Text\\
         S7~\cite{Medeiros:2019:atoms} & Atoms of confusion & C-like pseudo & 97 & 12 & 1164 & Comparison&1164 & Likert Scale\\
         \multirow[t]{3}{*}{S8~\cite{Siegmund2017}} & \multirow[t]{3}{30mm}{Identifier naming and layout}& \multirow[t]{3}{*}{Java} & \multirow[t]{3}{*}{14} & \multirow[t]{3}{*}{60} & \multirow[t]{3}{*}{420} & Similarity& 336 & Yes/No\\
          & & & & & &Output&42 & Yes/No\\
          & & & & & &Find syntax errors&42 & Yes/No\\
         \multirow[t]{2}{*}{S9~\cite{Peitek:2018:simultaneous}} & \multirow[t]{2}{30mm}{Task comparison} & \multirow[t]{2}{*}{Java} & \multirow[t]{2}{*}{22} & \multirow[t]{2}{*}{25} & \multirow[t]{2}{*}{550} & Output&440 & Yes/No\\
         & &  & &  &   &Find syntax errors&110 & Count\\
         \bottomrule
    \end{tabular}
\end{table*}
\subsection{Data Preparation and Prompt Construction}

\begin{figure}
    \centering
    \PromptBox{You will be shown a series of <Language> code snippets, which you should <Topic>.\\
    For each snippet you should respond by <Task>.\\
    <Other Info>.
    }
    \caption{Standardized template for task descriptions in prompts. Placeholders in $<$ $>$ are replaced with information relevant to that study. Other Info is used to communicate any additional information necessary for the study (e.g., response modalities, task variants).}
    \label{fig:template}
\end{figure}

\begin{figure*}
    \centering
    \footnotesize{
    \begin{align*}
    \Delta(\{\text{\SngCodeNoDesc}\}) &= d(\text{\SequentialNoCode}) - d(\text{\SequentialCode})\\
    \Delta(\{\text{\textsc{Desc}}\}) &= d(\text{\SeqCodeNoDesc}) - d(\text{\SequentialCode})\\
    \Delta(\{\text{\textsc{Sequ}}\}) &= d(\text{\SingleCode}) -d(\text{\SequentialCode})\\
    \Delta(\{\text{\textsc{Sequ, Desc}}\}) &= (d(\text{\SingleNoCode}) - d(\text{\SequentialCode})) - (\Delta(\textsc{\{Sequ\}}) + \Delta\text{(\{\SngCodeNoDesc\})})\\
    \Delta(\text{\textsc{\{Sequ, Code\}}}) &= (d(\text{\SngCodeNoDesc}) - d(\text{\SequentialCode})) - (\Delta(\{\text{\SeqNoCodeNoDesc}\}) + \Delta\text{(\{\SingleNoCode\})})\\
    \Delta(\textsc{\{Code, Desc\}}) &= (d(\text{\SeqNoCodeNoDesc}) - d(\text{\SequentialCode})) - (\Delta(\{\text{\SngCodeNoDesc}\}) + \Delta\text{(\{\SingleNoCode\})})\\
    \Delta(\textsc{\{Sequ, Code, Desc\}}) &= (d(\text{\NoInformation}) - d(\text{\SequentialCode}))\\
    &- (\Delta(\{\text{\SngCodeNoDesc}\}) + \Delta(\{\text{\SingleNoCode}\}) + \Delta(\{\text{\SeqNoCodeNoDesc}\}))\\
    &-(\Delta(\{\textsc{Sequ, Desc}\}) - \Delta(\textsc{\{Sequ, Code\}}) - \Delta(\textsc{\{Code, Desc\}}))
    \end{align*}
    }
    \caption{Formulas used to calculate the effects and interactions of removing sources of information. The effect size for an ablation $x$ is denoted as $d(x)$. The effect/interaction of removing a set of sources of information $X$ is denoted as $\Delta(X)$. }
    \label{fig:formulas}
\end{figure*}
After collecting the datasets, we translated all studies into natural language prompts to enable evaluation with Centaur and Llama. 
For the general layout of the prompts, we followed the prompt format used by Binz et al.~\cite{Binz:2025:Centaur} in training Centaur. 

Each prompt begins with a short description of the tasks in the study.
Where possible, we closely followed the original instructions provided to participants, omitting only task-irrelevant aspects (e.g., informed consent or compensation). 
Because for some studies original instructions were unavailable, we created a standardized template (see \autoref{fig:template}) covering aspects commonly disclosed in experiment instructions. 
The remainder of each prompt consists of tasks as seen by participants. 
Across studies, tasks generally follow a similar scheme: The participant is first shown one or more code snippets, answers questions about them, and proceeds sequentially. Studies differ primarily in the type of snippets used and in what kind of tasks participants are asked to perform (see~\autoref{tab:IncludedExperiments}).

To ensure compatibility with Centaur and comparability across datasets, we applied minor formatting transformations that preserve the original semantics. This included resolving tokenization issues and normalizing response options. We provide all details and examples in our replication package~\cite{zenodo:dataset}.

For RQ\textsubscript{1} we constructed one prompt per participant, where each prompt contains the task description, and the exact sequence of tasks and code the participant received in the study together with the responses they gave for these tasks.

To create the ablations used in RQ\textsubscript{2}, we first separated the prompt into three primary sources of information:
1) the task description, questions, and response options (\SingleNoCode); 2) the code used in tasks (\SngCodeNoDesc); and 3) information about preceding tasks and the responses of the participant (\SeqNoCodeNoDesc).
In the ablations, we systemically vary the availability of these three sources of information. 
Removing multiple sources might result in interaction effects, for example if both task description and prior responses indicate the response options for the next task independently of each other (e.g., in a multiple choice task).
In this case, we would see an interaction as the response options are still inferrable when removing either \SingleNoCode{} or \SeqNoCodeNoDesc{}, but removing both would make it impossible to infer the response options, resulting in a larger performance degradation than expected based on individual removal.
To enable us to detect interaction effects, we consider all possible combinations of the three sources, yielding 8 different categories of prompts. 

We refer to each category based on the sources of information present in the prompts, so the full prompts used in RQ\textsubscript{1} are referred to as \SequentialCode. We use \NoInformation~to refer to prompts with no sources of information present. We provide a full overview of all eight prompt categories and the sources of information found in each in \autoref{fig:method_overview_full}.

As in Binz al.~\cite{Binz:2025:Centaur}, we provide Centaur and Llama with the prompts as input and retrieve the next-token predictions for every token in the prompt. These predictions assign each token in the vocabulary of the model a probability based on all prior tokens in the prompt.  From these, we then collect the next-token predictions for each token position inside a response.

For this reason, we create one prompt per participant for all prompt categories that include \SeqNoCodeNoDesc{}, covering all tasks seen by that participant and the responses they gave for these tasks. Thus, for predicting the response to the $n$-th trial, the model has access to the task description, the participant's previous $n-1$ trials and responses, and the $n$-th task without its response. In the ablation scenarios of RQ\textsubscript{2}, the model has access to the same information except for the information removed by the ablations. If a response consists of multiple tokens, the predictions for each token in the response will be based on all prior tokens including any tokens that are part of the response, essentially representing conditional probabilities based on the probability of selecting the correct token for each prior token in the response.

For all categories without \SeqNoCodeNoDesc{} information, our ground truth is the set of all participant responses, so there is no single expected response. To enable the evaluation of multi-token responses, we create one prompt for each unique set of responses given to a specific task, covering only exactly that task and the relevant responses.

\subsection{Evaluation}

As in Binz al.~\cite{Binz:2025:Centaur}, we evaluate the performance of each model based on the negative log likelihood (NLL) of that model predicting exactly the responses given by human participants.
For every token that is part of a response,  we retrieve the probability distributions of the next-token prediction and calculate the negative log likelihood of the model predicting the response token correctly.
To obtain the negative log likelihood for a given response, we sum the negative log likelihood for each token in the response. 
If a task contains multiple logically separated answers (e.g., different gaps in a cloze task), we treat each as a separate response and obtain negative log likelihood measurements for each response individually.
The resulting negative log likelihood quantifies how unlikely the observed human response is according to the model, with lower values indicating better alignment.

For evaluating the performance of a model for prompt sets with \SeqNoCodeNoDesc{} information, we obtain a negative log likelihood measurement for each participant response in the dataset. 

For prompt sets with no \SeqNoCodeNoDesc{} information present, we initially obtain one measurement of negative log likelihood for each unique response to each task. To obtain a data point per participant response, we duplicate each data point to match the number of times the corresponding response was given across participants.

To answer RQ\textsubscript{1}, we compare the performance of Centaur and Llama across all studies and within each study individually using two-sided paired Student's t-tests~\cite{Student1908} to assess whether their mean performances differ significantly. 
We additionally report effect sizes using Cohen's d~\cite{Cohen1977} ($d(\text{X})$ represents the effect size for prompt variant X). 

To answer RQ\textsubscript{2}, we first calculate the same statistics as in RQ\textsubscript{1} for all seven categories of ablated prompts.
We then assess the impact of different sources of information by comparing effect sizes across prompt sets.
We use \SequentialCode~as a baseline and evaluate how removing different sources of information changes the difference in behavior between the two models. 
To calculate the individual effect differences of each source ($\Delta(\{\text{\textsc{Desc}}\})$, $\Delta(\{\text{\textsc{Code}}\})$, and $\Delta(\{\text{\textsc{Sequ}}\})$), we subtract the baseline effect size from the effect sizes of the ablated prompts missing just this information (\SeqCodeNoDesc, \SequentialNoCode, and \SingleCode).

To capture interactions between two sources, we first determine the total effect of removing both sources by calculating the difference between the ablation with both sources removed and the baseline. We then subtract the effects of removing both sources individually from that difference to obtain the difference in effect size caused by the interaction of the sources and separate it from the impact of the individual sources' effect differences.

Likewise, we calculate the interaction effect of removing all sources by calculating the difference between \NoInformation{} and \SequentialCode{} and subtracting the effect differences of all individual sources and pairwise interactions. Formulas for the calculation of all these effects and interactions can be found in \autoref{fig:formulas}.
A positive change in effect size when removing information indicates that the removed information was more relevant for Centaur (because in comparison Llama is less affected by removing it), whereas negative changes indicate greater relevance of the information for Llama.

\section{Results}

In this section, we present the results of our experiment structured along our two research questions.

\subsection{Performance of Centaur and Llama (RQ\textsubscript{1})}

\begin{table}
    \centering
    \caption{Performance comparison between Centaur and Llama across studies (RQ\textsubscript{1})}
    \begin{tabular}{lllrrrr}
        \toprule
        Study  & \multicolumn{2}{c}{NLL (Mean $\pm$ SD)}  & $t$ & $p$ value & Effect\\
       & Centaur & Llama &  & (2~sided) & Size d\\
        \midrule
S1~\cite{Ajami:2019:complexity} & $0.81 \pm 1.40$ & $0.89 \pm 1.79$ & $-4.09$ & $<0.001$ & $-0.09$ \\
S2~\cite{Bauer:2019:Indentation} & $4.40 \pm 2.97$ & $5.41 \pm 4.43$ & $-5.65$ & $<0.001$ & $-0.60$ \\
S3~\cite{Borstler:2016:chains} & $4.27 \pm 6.25$ & $4.87 \pm 7.10$ & $-10.22$ & $<0.001$ & $-0.20$ \\
S4~\cite{Buse:2010:metricreadability} & $1.40 \pm 0.72$ & $1.56 \pm 1.07$ & $-25.22$ & $<0.001$ & $-0.23$ \\
S5~\cite{Gopstein:2017:atoms} & $1.32 \pm 2.18$ & $1.55 \pm 2.81$ & $-14.05$ & $<0.001$ & $-0.18$ \\
S6~\cite{Gopstein:2017:atoms} & $4.16 \pm 3.35$ & $4.24 \pm 3.50$ & $-1.12$ & $0.26$ & $-0.05$ \\
S7~\cite{Medeiros:2019:atoms} & $1.62 \pm 0.69$ & $1.56 \pm 1.07$ & \textbf{$+2.16$} & $0.03$ & \textbf{$+0.06$} \\
S8~\cite{Siegmund2017} & $0.46 \pm 0.93$ & $0.62 \pm 1.45$ & $-4.53$ & $<0.001$ & $-0.22$ \\
S9~\cite{Peitek:2018:simultaneous} & $0.53 \pm 0.87$ & $0.66 \pm 1.33$ & $-4.89$ & $<0.001$ & $-0.21$ \\
         \midrule
Overall & $1.63 \pm 2.48$ & $1.85 \pm 2.96$ & $-29.46$ & $<0.001$ & $-0.17$ \\
        \bottomrule
    \end{tabular}
    \label{tab:rq1}
\end{table}
To evaluate the performance of Llama and Centaur in simulating program comprehension behavior, we first start with a general overview of the data, then compare the performance using statistical tests.

\paragraph{Overview of Performance}

Over all trials of all studies, Centaur achieves an average negative log likelihood of $1.63$, representing an improvement over Llama, which achieves an average negative log likelihood of $1.85$.
However, the performance of both Centaur and Llama differs significantly across datasets, with average negative log likelihoods of human responses varying between $0.62$ and $5.41$ for Llama and $0.46$ and $4.40$ for Centaur.
The most notable difference exists between studies with longer free text responses and those without such responses. 
The three studies with such tasks show very high negative log likelihoods for both models ($>4.0$ in all cases), while the other tasks show significantly lower negative log likelihoods ($<1.7$ in all cases). 

\paragraph{Evaluation}

The improvement of Centaur over Llama is statistically highly significant ($p<0.001$) with an effect size with a magnitude of $0.172$.
Centaur significantly outperforms Llama for 7 out of 9 studies, 
Of these, two show negligible effect sizes, four small effect sizes and one a medium effect size.  
Of the remaining two studies, one shows no significant difference and the other shows a significant, albeit negligible, decrease in performance for Centaur over Llama.
We provide detailed results for individual studies in~\autoref{tab:rq1}.

\RQAnswer{RQ\textsubscript{1}}{Compared to Llama, Centaur produces outputs significantly closer to human responses in a program comprehension setting both overall and in 7 out of 9 investigated studies.}

\subsection{Explanation of Performance Impact via Ablations (RQ\textsubscript{2})}
\label{sec:results_RQ2}

We performed 7 ablations that vary the amount of information included in the prompt and analyze the change in effect size caused by these ablations. We first provide an overview of the model performance in the ablations and then present the calculated source and interaction effects.

\begin{figure}
    \centering
    \includegraphics[width=0.9\linewidth]{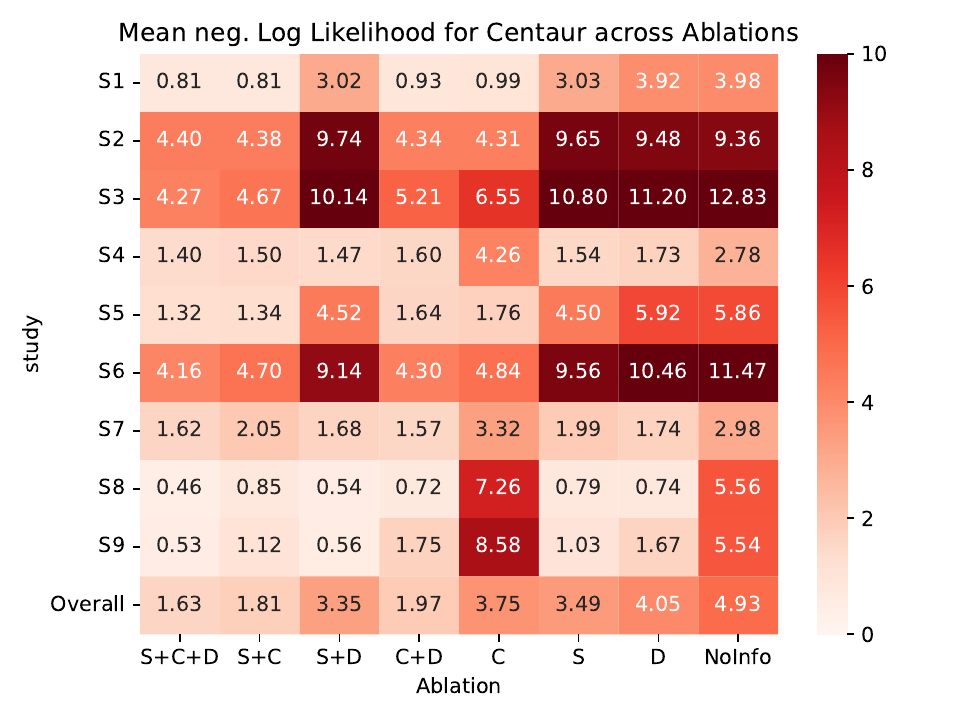}
    \includegraphics[width=0.9\linewidth]{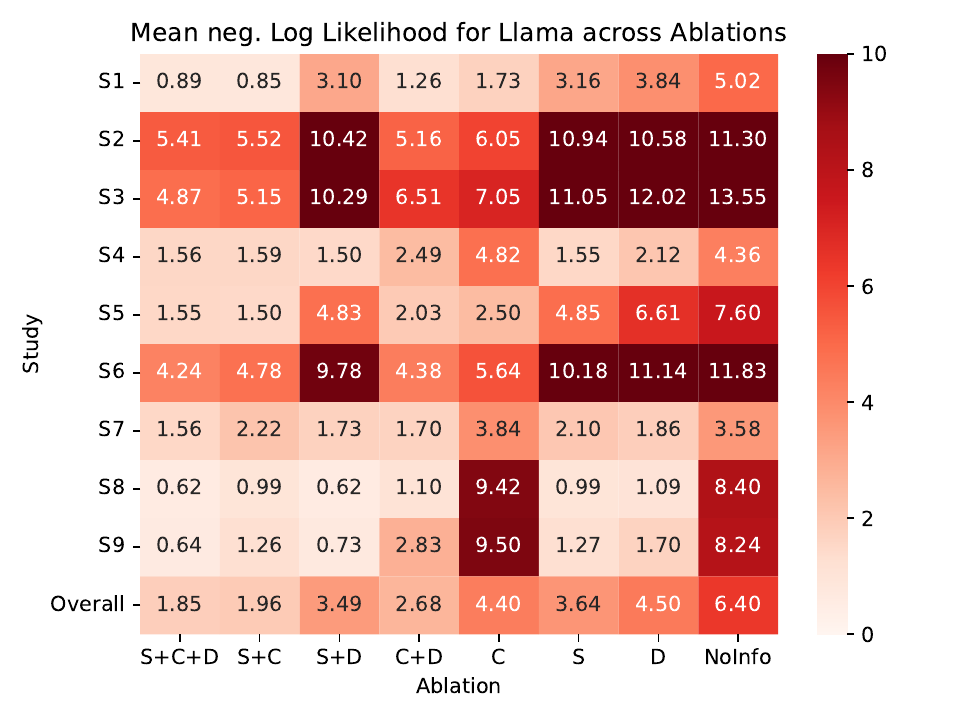}
    \vskip -1ex
    \caption{Heatmaps of the mean log likelihood across all studies and ablations for Centaur (top) and Llama (bottom). Sources of information are abbreviated to their initial letter for
readability}
    \label{fig:ablations_means}
\end{figure}

\paragraph{Overview of Performance}

Comparing the ablations to \SequentialCode{} shows that removing information reduces performance for both models in almost all cases~($58/63$).
When comparing the performance between models, Centaur outperforms Llama in all but one ablation (ablation \SingleNoCode{} for S1).
Across ablations, the studies can be clustered into two clusters with quite homogeneous trends in performance.

The first cluster consists of studies S1, S2, S3, S5, and S6. For these studies, any ablation that removes \SngCodeNoDesc{} results in a dramatic decrease in performance, whereas removing \SingleNoCode{} or \SeqNoCodeNoDesc{} has little to no impact. The worst performance for both models is achieved most often when removing all sources of information (\NoInformation).

The second cluster consists of S4, S7, S8, and S9, for which removing any of the individual sources of information has relatively little effect, and that show the worst performance when both \SeqNoCodeNoDesc{} and \SingleNoCode{} are removed (\SngCodeNoDesc). 

We provide an overview of the mean negative log likelihoods across studies and ablations in \autoref{fig:ablations_means}.

\paragraph{Evaluation}

\begin{figure}
    \centering
    \includegraphics[width=0.9\linewidth]{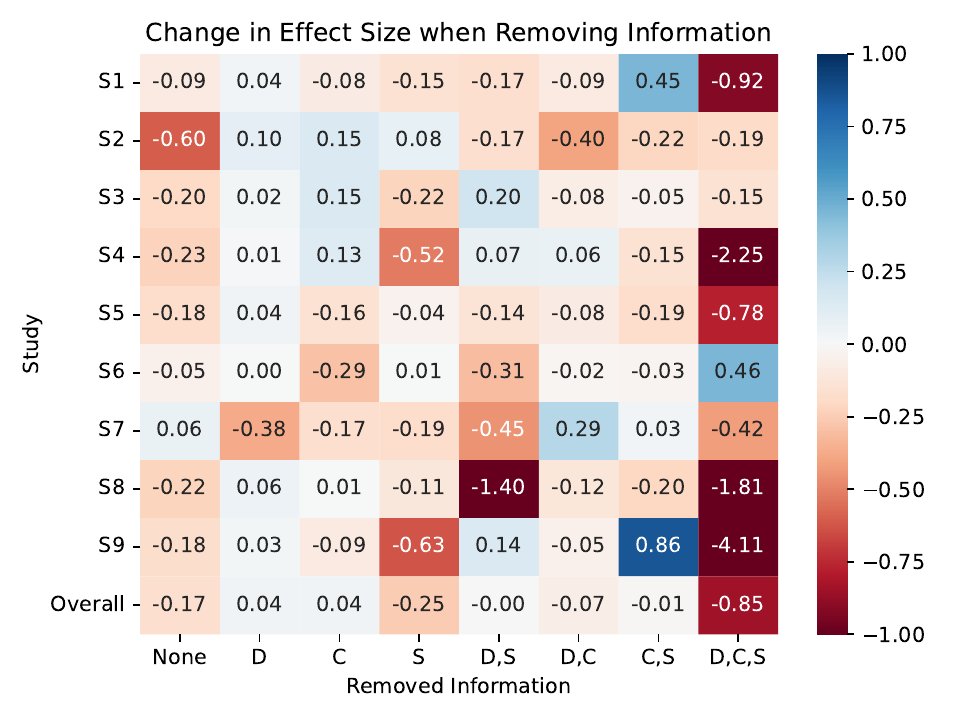}
    \vskip -1ex
    \caption{Heatmap of the effect of removing sources of information and their interaction effects for all studies. The leftmost column shows the effect size for a study when all information is present. Sources of information are abbreviated to their initial letter for readability.}
    \label{fig:ablations_change}
\end{figure}

When comparing both models across all trials, we find that removing \SeqNoCodeNoDesc{} information has a greater impact on the performance of Llama than Centaur, with the effect size increasing in magnitude to $-0.425$ (compared to $-0.172$ for the base version). 
On the other hand, removing \SngCodeNoDesc{} information appears to have a greater influence on Centaur, with the effect size slightly reducing in magnitude to $-0.127$. 
Similarly, \SingleNoCode{} information is overall more important for Centaur than for Llama, with the effect size reducing in magnitude to $-0.131$ when removing it.

Notably, these effects are weakened by an interaction when both \SngCodeNoDesc{} and \SingleNoCode{} are removed. 
Evaluated across all trials, the other two first-order interactions show only negligible effects. 
Removing all three sources of information sharply increases the effect size, indicating that Centaur's output with no relevant information is much closer to human responses than that of Llama. 

These findings indicate that the overall improvement of Centaur compared to Llama stems primarily from two sources, a better basic alignment with the expected response formats and an increase in benefit from task descriptions and code snippets compared to Llama, which benefits more from information about prior tasks being present. 

When investigating the effects of removing information in individual studies, we find some differences between studies. 

The effect of \SngCodeNoDesc{} information appears split between studies, with Centaur benefiting more from its presence in S2, S3, and S4, while Llama benefits more in S5, S6, and S7. 
The remaining three studies show far weaker effects in either direction.

The effect of removing \SeqNoCodeNoDesc{} differs primarily in magnitude, with Llama being more reliant on order information than Centaur in all studies except S2 and S6, with the strongest effect existing in S4 and S9.

For \SingleNoCode{}, we observe a consistent small effect across almost all studies indicating that Centaur benefits more from its presence then Llama.
The only notable exception is S7, for with a negative change in effect size occurs when removing \SingleNoCode{}, indicating that for this study Llama makes better use of the task description than Centaur. 
Notably, S7 is also the only study for which Centaur is outperformed by Llama in \SequentialCode. 

We provide an overview of all effects and interactions of removing sets of information in \autoref{fig:ablations_change}.

\RQAnswer{RQ\textsubscript{2}}{Overall, we find that improvements in performance of Centaur over Llama are primarily grounded in task descriptions and code snippets, with Llama relying more on response biases than Centaur in most datasets.}

\section{Discussion}

In this section, we discuss our results and their implications.

\subsection{Research Questions}

Since the early days of software engineering research, modeling and predicting program comprehension has been an ambitious goal. 
A range of theoretical accounts was proposed to explain how developers understand code, but these efforts largely stalled by the end of the 1990s~\cite{Wyrich:2023:CC}.
The core difficulty was not a lack of ideas but the inherent challenge of capturing a complex cognitive process in an interpretable and empirically testable form.
As a result, the field gradually shifted away from holistic theories toward more limited proxies, most notably complexity metrics and highly controlled laboratory studies isolating individual program features~\cite{Scalabrino:2019:Metrics,Wyrich:2023:40Years}.
While this work has produced valuable insights, it has also led to a fragmented understanding: Many findings are context-dependent and do not reliably generalize across tasks, while traditional metrics often show only weak relationships with actual comprehension behavior~\cite{Scalabrino:2019:Metrics,Peitek:2021:Metrics}.
In that sense, the field's ``big-picture'' model of program comprehension has changed surprisingly little for more than two decades.

Over 25 years later, our results presented here suggest that it might be possible to predict program comprehension behavior at scale using foundation models of human cognition.
In particular, our results in RQ\textsubscript{1} show that behavioral patterns learned by Centaur through fine-tuning on psychological experiment data transfer to diverse program comprehension tasks.
Importantly, RQ\textsubscript{2} further demonstrates that this effect is not merely driven by response biases but reflects improved use of task-relevant information such as code and natural language task descriptions compared to Llama.
Our findings represent a meaningful shift in perspective: Rather than treating program comprehension as a set of isolated phenomena, our results are consistent with the view that it reflects the interaction of fundamental cognitive processes across domains.
At the same time, the breadth of psychological training data underlying Centaur leaves an important question open: Whether the observed transfer occurs due to cognitive processes specific to some activities or broader regularities that occur across different experiments.
Addressing this question could provide a more fine-grained mapping between cognitive processes and specific aspects of program comprehension, thereby helping reconstruct a more structured theory at the cognitive level.

A key distinction to existing approaches is that Centaur is not an interpretable model of cognition in the traditional sense.
However, it generalizes far better across tasks than existing approaches and, unlike existing models, can predict human behavior reasonably well.
This creates an interesting methodological complementarity: While interpretability has historically been central in program comprehension research, predictive models such as Centaur can serve as a complementary tool for hypothesis generation and theory refinement.
Similar ideas have recently been explored in psychology, where Centaur has already been used to support the development of more interpretable white-box models of human behavior~\cite{Binz:2025:Automated}.
In this sense, predictive foundation models may serve as a bridge toward more structured and eventually interpretable theories of program comprehension.

Beyond theoretical implications, the results also point toward potential practical applications.
The observed improvements in predictive performance suggest that foundation models of human cognition could be used to estimate when code is likely to be difficult to understand.
Compared to traditional metrics, such models have the advantage of being trained on behavioral data and therefore implicitly incorporating factors that are typically ignored in static metrics, such as naming or code-level documentation.
Given the increasing prevalence of AI coding assistants, the role of developers is shifting toward understanding and validation of generated code~\cite{Felder:2026:Adoption,Vella:2026:Impactaicodingassistants}, making accurate predictions of comprehension difficulty increasingly relevant for real-world workflows.

\subsection{Robustness and Variability Across Studies}

Beyond the overall transfer effects observed in RQ\textsubscript{1} and RQ\textsubscript{2}, our results reveal several systematic differences in how Centaur and Llama respond across studies and ablation conditions.

The most notable difference between studies occurs in the \NoInformation{} ablation (i.e., when removing all sources of information). For studies S1, S4, S5, S7, S8, and S9, we observe a strong interaction indicating that Centaur is substantially less affected by removing all information than Llama.
A plausible explanation lies in the response structure of these studies: In all six cases, responses are predominantly single-character outputs.
As Centaur has been fine-tuned on a large number of experiments involving similarly constrained response formats, it is more likely to respond with a token representing a single character, even in the absence of informative input, whereas Llama shows a stronger dependency on explicit task cues.

A second set of differences emerges in the \SequentialNoCode~ablations, with Llama making better use of the removed \SngCodeNoDesc~information for S5, S6, and S7, while Centaur makes better use of it for S2, S3, and S4, with the remaining three studies showing much weaker effects in either direction. 
Notably, S5, S6, and S7 all concern \emph{Atoms of Confusion}~\cite{Gopstein:2017:atoms}, i.e., syntactic code patterns known to induce misunderstandings in human readers.
One possible interpretation is that these patterns engage cognitive processes that differ qualitatively from standard code comprehension and that these processes are less well represented in Centaur's training data.
This suggests a potential boundary condition for transfer, where certain classes of cognitively demanding constructs are not captured equally well by the learned regularities.
It is possible that this boundary could be overcome by extending the fine-tuning of the model with data from tasks involving such constructs, as this would allow the model to learn about how behavior might differ in such tasks.  

Finally, when comparing the results of RQ\textsubscript{1} to those reported in the original Centaur article by Binz et al.~\cite{Binz:2025:Centaur}, we observe higher variability in the predictive performance for program comprehension tasks, as reflected in the standard error of the mean negative log likelihood.
This increased variance is plausibly driven by heterogeneity in participant expertise: Unlike many controlled psychological experiments, program comprehension studies often include both novice and expert developers, whose cognitive strategies and performance levels differ substantially~\cite{Busjahn2015, Peitek2022, Latoza:2007:factfinding}.
This mixture likely introduces additional variability that is not present in more homogeneous experimental settings.

\subsection{Perspective}

Our findings open several directions for advancing both the theoretical understanding and practical use of foundation models of human cognition in software engineering.

From a theoretical perspective, a central open question concerns the nature of the transfer observed in this work.
While our results show that behavioral regularities learned from psychological experiments generalize to program comprehension, it remains unclear which parts of this training signal are most relevant.
A promising direction is to systematically analyze the contribution of different types of psychological experiments, for instance, by ablating subsets of the training data used for Centaur.
Such analysis could reveal whether specific cognitive processes are particularly predictive of program comprehension behavior or whether the transfer is driven by more general regularities.
This, in turn, could provide a more fine-grained and interpretable link between cognitive processes and program comprehension, helping to move beyond purely predictive models toward more refined theories.

From a practical perspective, improving predictive performance remains an important goal.
One natural extension is to further fine-tune models like Centaur on program comprehension datasets directly.
Incorporating domain-specific behavioral data could enhance accuracy and robustness, particularly for phenomena that appear underrepresented in the original training data.
Put generally, it would be worthwhile to attempt to develop a hybrid training paradigm in which general cognitive regularities are complemented by targeted domain data.

Finally, our results raise the question of how far this approach can generalize within software engineering.
Program comprehension is a central but not unique cognitive task in this domain.
Other activities, such as debugging or reviewing pull requests, share similar levels of cognitive complexity and may also be suitable for prediction through models of human behavior.
Exploring such applications could significantly broaden the applicability of this line of work and provide novel tools for studying developer behavior across a wide range of software engineering tasks.

\section{Threats to Validity}

This section details the trade-offs we made as well as the limitations of our methodology and findings.

\subsection{Construct Validity}

Our evaluation of the capabilities of Centaur in program-comprehension settings is limited to the prediction of behavioral responses, since that is the task Centaur has been trained on. 
However, there are other aspects, such as task duration and neurophysiological measurements, that are commonly used in experiments to evaluate program comprehension. 
Our findings do not allow us to make claims regarding the efficacy of using foundation models of human cognition to predict these aspects. 
Future work shall comprehensively target program comprehension for simulation with foundation models trained for other aspects, such as brain activation~\cite{Dascoli2026} or vision~\cite{Zanca2025}.

We measure how well models simulate behavioral data from a collection of experiments to operationalize how well they model human behavior in program comprehension. 
As such, biases and limitations present in these studies may limit to what extent we capture program comprehension behavior.
We reduce the effect of biases and limitations of individual studies on our evaluation by comparing Centaur and Llama across multiple studies. 
However, effects present in multiple investigated studies might still affect the validity of our results.

\subsection{Internal Validity}

To enable an evaluation with Centaur, we translated experiments into natural language. 
This process required a number of choices (e.g., wording of the task description), which could influence the behavior of the investigated models in these experiments. 
To limit this influence we followed real task descriptions as closely as possible and otherwise used consistent approaches across studies.

We assume that changes in effect size after removing specific information are caused by differences in how the models process this information. 
Yet, given the black-box nature of the models we investigated, it is possible that removing the information had effects that were not related to the information itself, such as due to changing the size of the prompts.

\subsection{External Validity}

Our analysis focuses on Centaur and Llama~3.1. In line with Binz et al.~\cite{Binz:2025:Centaur}, we used the non-fine-tuned version of Llama~3.1 as baseline comparison to analyze the impact of fine-tuning models on psychological data. 
While this design allows us to isolate the effect of the fine-tuning on behavioral data, it is possible that our results may not generalize to other foundation models of human cognition built on more advanced LLMs than Llama~3.1. 

Another threat to external validity arises from the dataset selection. To reduce bias, we used a comprehensive secondary study~\cite{Wyrich:2023:40Years} to identify experiments concerning program comprehension and select datasets. Unfortunately, more than 75\% of the experiments did not provide sufficient replication packages or any at all. Moreover, the selected studies primarily consisted of program-comprehension tasks in a lab setting with comparatively small code snippets. As a result, our evaluation is limited to nine specific datasets, which provides evidence for transfer to a diverse set of program-comprehension tasks but may not fully capture the entire breadth of program comprehension.

In the same vein, we excluded studies if a textual representation was not possible or if their responses were insufficiently constrained for an evaluation based on exact matches (i.e. longer free-text responses). 
Similarly, we omitted all trials in which meta events, such as participants aborting a study, would have to be predicted, since Centaur was not trained to predict such circumstances. 
While these exclusions allowed us to maximize internal validity regarding our goal of evaluating the performance of Centaur, they may introduce a bias towards specific aspects of program comprehension in our sample.

\section{Conclusion}
\label{sec:conclusion}

In this paper, we investigated whether foundation models of human cognition can predict human behavior in program-comprehension tasks. 
Across nine datasets covering different kinds of program-comprehension tasks, we found that Centaur predicts human behavior in program-comprehension tasks significantly better than its base model, Llama 3.1.
Our ablation studies revealed that this performance improvement occurs not due to generic human response biases but because Centaur makes better use of task-relevant information, such as task descriptions and code snippets. 

Our results provide first empirical evidence that foundation models of human cognition are able to learn domain-general behavioral patterns that generalize to the more complex setting of program comprehension, despite the absence of such tasks in the training data.
From a practical perspective, foundation models of human cognition offer a promising complement to traditional complexity metrics by enabling behavior-aware predictions of how developers engage with source code.
By jointly considering code structure, task context, and learned behavioral regularities, they may provide more targeted predictions of real-world comprehension difficulty.
From a research perspective, this supports the view of program comprehension as a complex activity grounded in fundamental, domain-general cognitive processes and represents a big step towards a more unified understanding of developer behavior. 
In this light, foundation models of human cognition offer a promising new perspective for studying and predicting how developers interact with code and other artifacts.

\section*{Data Availability}
\label{sec:dataavailability}
Following open science principles in software engineering~\cite{Mendez:2020:Open}, we openly disclose details on our dataset search, specific prompts for all analyzed studies, scripts (including details on how we modified prompts to address tokenization issues), and raw data~\cite{zenodo:dataset}.

\section*{Acknowledgment}
% ANON
This work has been supported by the European Union as part of ERC Advanced Grant ``Brains On Code'' (101052182).

\bibliographystyle{IEEEtran}
\bibliography{main}

\end{document}